\documentstyle[psfig,pre,multicol,aps]{revtex}

\begin{document}
\title{\bf Localization of nonlinear  excitations in curved waveguides}
\author{Yu. B. Gaididei}
\address{Bogolyubov Institute for Theoretical Physics,
Metrologichna str. 14 B, 01413, Kiev, Ukraine}
\author{   P.L. Christiansen}
\address{Informatics and Mathematical Modelling, 
The Technical University of Denmark, 
DK-2800 Lyngby, Denmark}
\author{P. G. Kevrekidis} 
\address{Department of Mathematics and Statistics,
University of Massachusetts, Amherst, MA 01003-4515, USA}
\author{H. B{\"u}ttner} 
\address{Physikalisches Institut, Universit\"{a}t
Bayreuth, Bayreuth D-95440, Germany}
\author{A.R. Bishop} 
\address{
\date{\today}Theoretical Division and Center for 
Nonlinear Studies,Los Alamos National Laboratory,
Los Alamos, NM 87545, USA}
\maketitle

\begin{abstract}
Motivated by the example of a curved
waveguide embedded in a photonic crystal, we examine
the effects of geometry in a ``quantum channel'' of parabolic
form. We study the linear case and derive exact as well as
approximate expressions for the eigenvalues and eigenfunctions
of the linear problem. We then proceed to the nonlinear
setting and its stationary states in a number of limiting cases
that allow for analytical treatment. The results of our analysis
are used as initial conditions in direct numerical simulations
of the nonlinear problem and localized excitations are found
to persist, as well as to have interesting relaxational dynamics.
Analogies of the present problem in
contexts related to atomic physics and particularly to
Bose-Einstein condensation are discussed. 
\end{abstract}

\section{Introduction}
Localization phenomena are widely recognized as  key to understanding  the excitation 
dynamics  in many physical contexts such as light propagation, charge and energy transport 
in condensed-matter physics and biophysics, and   
Bose-Einstein condensation of dilute atomic gases \cite{christ,souk}.
Recent advances in micro-structuring technology have made it possible to 
fabricate various low-dimensional systems with complicated geometry. 
Examples are photonic crystals with embedded  defect structures such as 
microcavities, waveguides  and waveguide bends \cite{souk,mekis1,noda,kkt}, 
narrow
structures (quantum dots and channels) formed at semiconductor 
heterostructures \cite{wees,reed,ismail}, 
magnetic nanodisks, dots  and rings \cite{Pulwey01,shinjo,klaeui}.  

On the other hand, it is well known that the wave equation  subject to
Dirichlet boundary conditions  has  bound states  in straight 
channels of variable width
\cite{schult,berggren,andrews}, and in curved channels of 
constant cross-section \cite{exner,gold}.
 Spectral and transport characteristics of quantum electron channels 
\cite{vakh} and   
 waveguides in photonic crystal 
   \cite{mekis1} are  in essential ways  modified by the existence of segments with finite curvature. 
The two-dimensional Laplacian operator supported by an infinite curve 
  which is asymptotically
  straight has at least one bound state below the threshold of the continuum spectrum, as was 
  recently proved in \cite{exner01}. The appearance of an effective attractive potential 
  in the wave equation  is due to constraining quantum particles
   from higher to lower dimensional manifolds \cite{costa,jaffe,exnkr}. Curvature induced 
   bound-state 
energies and and corresponding wave functions were studied  in \cite{encinosa}.

  Until recently there have been few  theoretical and numerical studies of the effect 
  of curvature on properties of nonlinear excitations. Nonlinear whispering gallery modes
  for a nonlinear Maxwell equation in microdisks 
were investigated in  \cite{harayama};
  the excitation of whispering-gallery-type electromagnetic modes by a moving  fluxon in an
  annular Josephson junction was shown in \cite{ustinov}.  Nonlinear localized modes 
  in two-dimensional photonic crystal waveguides were studied in \cite{mingkivsh}.
  A curved chain of nonlinear oscillators  was considered in \cite{curve} and it was shown 
   that the interplay of curvature 
  and nonlinearity leads to a symmetry breaking when an
asymmetric stationary state becomes energetically more favorable than a symmetric 
stationary state. 
  Propagation of Bose-Einstein condensates in magnetic waveguides was 
experimentally demonstrated  
quite 
  recently in \cite{leanhardt}; single-mode propagation was observed along 
  homogeneous segments of the waveguide, while geometric deformations of the microfabricated
  wires  led to strong transverse excitations.

Motivated by the experimental relevance of the above mentioned 
geometric deformations, in the present work we   
 aim at investigating nonlinear excitations  in  a prototypical setup
incorporating such phenomena. As our case example, we will examine
an infinitely 
 narrow  curved nonlinear waveguide (channel) embedded  in 
two-dimensional linear medium  (idle region). 

Our presentation will proceed as follows:
in Section II, we set up the mathematical model of interest
and examine its general properties and equations of motion.
In Section III, we study the linear case and present its
explicit solutions for the bound states, as well as for the 
corresponding eigenvalues. In section IV, we focus on the
nonlinear problem, while in section V we supplement our
analysis with numerical results. Finally, in section VI, we 
summarize our findings and present our conclusions. 
The appendix presents some additional technical details on
the solution of the linear problem and its Green's function
formulation.

\section{System and Equations of Motion}
Our model is described by the Hamiltonian
\begin{eqnarray}\label{ham}
H=\int\limits_{-\infty}^{\infty}\{|\nabla\psi|^2-\nu\,(|\psi|^2-
A\frac{1}{2}|\psi|^4)\,\delta(y-f(x))\}dxdy,\end{eqnarray}
where $\psi(\vec{r},t)$ is the complex amplitude function,
$\vec{r}=(x,y)$, $\nabla^2=\partial_x^2+\partial_y^2$,
$\nu$ is the energy difference between the quantum channel and
the passive region (refractive index difference in the case of
photonic crystals and waveguides), 
the coefficient $A$  characterizes   the nonlinearity of the 
medium, e.g., the nonlinear corrections to the refractive index of the photonic band-gap 
materials, or self-interaction of the quasi-particles in the quantum channel.
The function $y=f(x)$ gives the shape of the channel. 
From the Hamiltonian, we obtain the equation of motion in the form
\begin{eqnarray}\label{eq}
i\partial_t \psi(\vec{r},t)+\nabla^2\psi+\nu \,\delta(y-f(x))
F(|\psi|^2)\,\psi=0,
\end{eqnarray}
where the function $F(|\psi|^2)$ is given by 
\begin{eqnarray}\label{refr}
F=1-A|\psi|^2.\end{eqnarray}
 Equation (\ref{eq}) has as integrals of motion the
Hamiltonian (\ref{ham})
and the ($L^2$) norm (referred to  e.g., as the 
number of atoms in BEC or power in 
nonlinear optics)
\begin{eqnarray}\label{norm}
N=\int\limits_{-\infty}^{\infty}|\psi|^2\,dxdy.\end{eqnarray}

It is well known 
(see {\it e.g.}, \cite{encyklop}) that if the edges of the Fr{\'e}net trihedron (the tangent, 
the principal normal and the binormal)
at a given point  are considered as the axes of a Cartesian coordinate system, then the equation
of the  curve in a neighbourhood of this point  has the form
\begin{eqnarray}\label{parametr}
x=s+\cdots,~y=\frac{\kappa}{2}\,s^2+\cdots,~z=-\frac{\kappa\,\tau}{6}\,s^3+\cdots
\end{eqnarray} 
where $s$ is the arclength, $\kappa$ is the curvature and $\tau$ is the torsion 
of the curve at this point. 
Thus if the curvature  of the plane curve is not too large one can represent it as a parabola
\begin{eqnarray}\label{parabola}
y=\frac{\kappa}{2}\,x^2.
\end{eqnarray}

In this case  Eq. (\ref{eq}) takes the form
\begin{eqnarray}\label{eqpar}
i\partial_t\psi+\nabla^2\psi+\nu \,\delta(y-\frac{x^2}{2R})F(|\psi|^2)\,\psi=0,
\end{eqnarray}
where $R=1/\kappa$ is the maximum radius of curvature of the curve. 
It is convenient to use the parabolic coordinates
\begin{eqnarray}\label{parcoor}
x=\frac{u v}{R},~~~y=\frac{R}{2}+\frac{1}{2 R}(u^2-v^2). \end{eqnarray}
The coordinate lines are two orthogonal families of confocal parabolas, 
with axes along the $y$ axis. These lines are given by
$$R \frac{ x^2}{v^2}=2y-R+v^2/R,~~~R \frac{x^2 }{u^2}=-2y+R+u^2/R$$
or \begin{eqnarray}\label{uv}
u/\sqrt{R}=\pm\sqrt{y-\frac{R}{2}+\sqrt{(y-\frac{R}{2})^2+x^2}},\nonumber\\
v/\sqrt{R}=\sqrt{-y+\frac{R}{2}+\sqrt{(y-\frac{R}{2})^2+x^2}}.\end{eqnarray}
The variable $u$ is allowed to range from $-\infty$ to $\infty$, whereas
$v$ is positive. Introducing Eqs. (\ref{parcoor}) into Eq. (\ref{eqpar})
and using the properties of parabolic coordinates (see e.g., \cite{djones}),
we obtain:
\begin{eqnarray}\label{eqparc}
i\,\frac{1}{R^2} (u^2+v^2)\,\partial_t\psi+\left(\partial_u^2+\partial_v^2\right)\psi+
\nu\,\delta(v-R)\,F(|\psi|^2)\,\psi=0. 
\end{eqnarray}

Using the Fourier transform with respect to $t$,
\begin{eqnarray}\label{fourier}
\bar{\psi}=\frac{1}{2\pi}\int\limits_{-\infty}^{\infty} e^{i\omega t} \psi(u,v,t)\,dt,
\end{eqnarray} where the bars denote the Fourier transformed quantities, one can represent Eq. (\ref{eqparc}) 
in the form
\begin{eqnarray}\label{eqfour}
-\frac{\omega}{R^2} (u^2+v^2)\,\bar{\psi}+\left(\partial_u^2+\partial_v^2\right)\bar{\psi}+
\nu\,\delta(v-R)\,\overline{F(|\psi|^2)\,\psi}(u,v)=0. 
\end{eqnarray}
 Equation (\ref{eqparc}) can, in turn, 
be expressed in the form of the integral equation
  \begin{eqnarray}
 \label{inteqn}
 \bar{\psi}(u,v,\omega)=\nu\, \int\limits_{-\infty}^{\infty}d u'\,
 \int\limits_{0}^{\infty}\,d v'\,G\left(u,v;u',v'\right)\,\delta(v'-R)\,\overline{
 F(|\psi|^2)\,\psi}(u',v',\omega),
 \end{eqnarray}
where the Green's function $G\left(u,v;u',v'\right)$ satisfies the equation

\begin{eqnarray}\label{eqgreen}
\left(\partial_u^2+\partial_v^2\right)\,G-\frac{\omega}{R^2} (u^2+v^2)\,G=-\delta(u-u')\,
\delta(v-v'), 
\end{eqnarray}
and has the form (see Appendix for details)
\begin{eqnarray}\label{green}
G\left(u,v;u',v'\right)=\frac{\sqrt{\pi}}{2}\sum_{n=0}^{\infty}\,F_n(u)\,F_n(u')\,
\{V_n(v)\,U_n(v')\,\theta(v'-v)+U_n(v)\,V_n(v')\,\theta(v-v')\}.
\end{eqnarray}
Here 
\begin{eqnarray}\label{hermite}
F_n(u)=a_n\,e^{-\sqrt{\omega} u^2/2 R}\,H_n\left(u\,\omega^{1/4}\,R^{-1/2}\right),~~~n=0,1,2,... ,
 \end{eqnarray}
 where 
 $H_n\left(z\right)$ is the Hermite polynomial \cite{abr}, $a_n=\left(\omega^{1/2}/R\,\pi\right)^{1/4}\,
 \left(2^n\,n!\right)^{-1/2}$ is the normalization constant, and 
  
\begin{eqnarray}\label{weber}
 V_{n}(v)= V\left(n+\frac{1}{2}, v\,\sqrt{2}\,\omega^{1/4}\,R^{-1/2}\right), 
 \nonumber\\
 U_{n}(v)=\,U\left(n+\frac{1}{2}, v\,\sqrt{2}\,\omega^{1/4}\,R^{-1/2}\right),
 \end{eqnarray}
 with $V(a,x)$ and $U(a,x)$  being the Weber parabolic 
 cylinder functions \cite{abr}. 
 
 It can then be 
seen from Eqs. (\ref{inteqn}) and (\ref{green}) that the Fourier transformed 
 channel wave function
 \begin{eqnarray}
\label{chanwf}\bar{\phi}(u,\omega)\equiv \bar{\psi}(u,v,\omega)\bigg |_{v=R},
\end{eqnarray}
 satisfies the equation
 \begin{eqnarray}\label{chaneq}\bar{\phi}(u,\omega)=\nu \,\omega^{-1/4}\,\frac{\sqrt{\pi\,R}}{2}\,
 \sum_{n=0}^{\infty}\,
 \int\limits_{-\infty}^{\infty}
 \,F_n(u)\,F_n(u')\,V_n(R)\,U_n(R)\,\overline{F(|\phi|^2)\,\phi}(u',\omega)\,d u', \end{eqnarray}
 
 or equivalently  the equation
 \begin{eqnarray}\label{chanequ}
 \nu \,\omega^{-1/4}\,\frac{\sqrt{\pi\,R}}{2}\,\,\overline{F(|\phi|^2)\,\phi}(u,\omega)=\sum_{n=0}^{\infty}\,
 \int\limits_{-\infty}^{\infty}
 \,\frac{F_n(u)\,F_n(u')}{V_n(R)\,U_n(R)}\bar{\phi}(u',\omega)\,d u'.
 \end{eqnarray}
 
 The wave function $\bar{\psi}(u,v,\omega)$ may be represented in terms of the channel 
 wave function $\bar{\phi}(u,\omega)$
 as follows:
  \begin{eqnarray}\label{totwf}
 \bar{\psi}(u,v,\omega)= \,\frac{\omega^{1/4}}{\sqrt{\,R}}
 \int\limits_{-\infty}^{\infty}\,d u'\,\bar{\phi}(u',\omega)\,\sum_{n=0}^{\infty}\,F_n(u)\,F_n(u')
 \,\frac{V_n(v)}{V_n(R)}~~~~~for ~~0<v<R,\nonumber\\
 \bar{\psi}(u,v,\omega)= \,\frac{\omega^{1/4}}{\sqrt{\,R}}
 \int\limits_{-\infty}^{\infty}\,d u'\,\bar{\phi}(u',\omega)\,\sum_{n=0}^{\infty}\,F_n(u)\,F_n(u')
 \,\frac{U_n(v)}{U_n(R)}~~~~~for ~~R<v.
 \end{eqnarray}
 
 \section{Linear case:$~A=0$}

 In the linear case ($A=0$),  Eq. (\ref{chanequ}) assumes the form
 \begin{eqnarray}\label{lineq}
 \nu \,\omega^{-1/4}\,\frac{\sqrt{\pi\,R}}{2}\,\,\overline{\phi}(u,\omega)=
 \sum_{l=0}^{\infty}\,
 \int\limits_{-\infty}^{\infty}
 \,\frac{F_l(u)\,F_l(u')}{V_l(R)\,U_l(R)}\bar{\phi}(u',\omega)\,d u'.
 \end{eqnarray}
Taking into account that the set of functions $F_n (u)$ is complete and orthonormal, 
 one can rewrite
 Eq.(\ref{lineq}) as
 \begin{eqnarray}\label{linequ}
 \sum_{l=0}^{\infty}\,\{\nu \,\omega^{-1/4}\,\frac{\sqrt{\pi\,R}}{2}
 \,U_l\left(R\right)\,
 V_l\left(R\right)-1\}\,\phi_l=
 0,
 \end{eqnarray}
 where \begin{eqnarray}\label{phiint} \phi_l= \int\limits_{-\infty}^{\infty}
 \,F_l(u)\,\overline{\phi}(u,\omega)\,du.\end{eqnarray}
 Thus, we can conclude that   the solution of the linear eigenvalue problem can be presented in the form
 \begin{eqnarray}\label{phieig} 
 \phi_l=\delta_{k n}\,{\cal N}_n \end{eqnarray}
 
 \begin{eqnarray}\label{lambda}
 \nu\,\sqrt{\frac{R}{\lambda_n}}
 \,U\left(n+\frac{1}{2}, \sqrt{2\lambda_n R}\right)\,
 V\left(n+\frac{1}{2}, \sqrt{2\lambda_n R}\right)\,=
 \frac{2}{\sqrt{\pi}}.
 \end{eqnarray}
 Eq. (\ref{lambda}) determines the  frequency   ($\omega_n=\lambda_n^2$) 
of the $n$-th  eigenstate and 
 Eq. (\ref{phieig}) with ${\cal N}_n$  being a normalization constant and $\delta_{k n}$ being the Kronecker delta, yields its amplitude.
 
 Introducing Eqs. (\ref{phiint}) and (\ref{phieig}) into Eqs. (\ref{totwf}) we obtain that the eigenfunction 
 $\Phi_n(u,v)\equiv  \bar{\psi}(u,v,\omega_n)$, which corresponds to the eigenvalue given by Eq. (\ref{lambda}), 
 can be expressed as:
\begin{eqnarray}\label{totwflin}
 \Phi_n(u,v)={\cal N}_n\,F_n(u)\,\left(\,\frac{V_n(v)}{V_n(R)}\theta(R-v)+\,\frac{U_n(v)}{U_n(R)}\,\theta(v-R)\right).
 \end{eqnarray}
 
    For even values of $n$:
 $n=2 m ~(m=0,1,2,..),  $ Eq. (\ref{lambda}) always has a solution and for 
 $\nu R \rightarrow 0$  and the eigenvalue is given by 
 \begin{eqnarray}\label{even}
 \lambda_{2m}\approx\left(\frac{\Gamma(m+\frac{1}{2})}{m!}\right)^2\,
 \frac{\nu^2 R}{4 (1+\nu R)^2}.
 \end{eqnarray}
  For  $n=2m+1$, $m=0,1,2,...$ the bound state exists only for
 $\nu R\geq 1$ and near the lower bound the energy of the bound state is 
 given by
 \begin{eqnarray}\label{odd}
 \lambda_{2m+1}\approx\left(\frac{m!}{2\,\Gamma(m+\frac{3}{2})}\right)^2\,
 \frac{(\nu\,R-1)^2}{\nu^2\,R^3}.
 \end{eqnarray} 
 In the limit of large radius of curvature $R$ and moderate $n$, {i.e.},
 $~\nu R\gg n+\frac{1}{2}~$, we obtain from 
 Eq. (\ref{lambda}) that the eigenvalues $\lambda_n$ are determined by
 \begin{eqnarray}\label{Rlarge}
 \lambda_n=\frac{\nu}{2}\,\left(1-\frac{2n+1}{\nu R}\right).
 \end{eqnarray}
 Thus the bound state energy  decreases when the curvature 
 of the chain increases 
and in the limit $R\rightarrow \infty$ 
we obtain the straight-line result: $\lambda=\nu/2$.
 
 It is interesting  to return to the Cartesian coordinates and 
 to consider the shape of the bound state wave function. Let us consider 
 the case $n=0$. It is seen from Eqs.  (\ref{solF}) and (\ref{solG})
  that
 \begin{eqnarray}\label{phi0}
 \Phi_0={\cal N}_0\,e^{-\lambda_0\,y}
 \left( {\rm erfc}\left(\sqrt{\lambda_0 R}\right)\theta\left(y-\frac{x^2}{2R}\right)+
 {\rm erfc}\left(v \sqrt{\lambda_0}\right)\theta\left(\frac{x^2}{2R}-y\right)\right),
 \end{eqnarray}
 where the function $v(x,y)$ is given by 
 Eq. (\ref{uv}) 
When  $R\rightarrow \infty$
 (straight waveguide) the wave function  is localized  in the $y$-direction only. However, for  finite $R$ 
 the function is localized both in $x$- and $y$-directions. The localization length in the $x$-direction
 is proportional to $R$. The expression of Eq. (\ref{phi0}) will also
be used as a starting point in our direct numerical simulations of  
Eq. (\ref{eqparc}); see section V below.
 
 \section{Nonlinear case}

In the following, 
we restrict ourselves to the case of  the  
waveguides with small or moderate curvature ($1/R < 1$).
 We use   Darwin's expansion of the parabolic cylinder functions \cite{abr}:
 \begin{eqnarray}\label{darwin}
 U(a,x)\,V(a,x)\approx \sqrt{\frac{2}{\pi\,(x^2+4a)}},~~~x^2+4a\gg 1.\end{eqnarray}
 Using this approximation, Eq. (\ref{chanequ}) may be represented in the form
  \begin{eqnarray}\label{chanequi}
 \nu \,\omega^{1/4}\,\frac{\sqrt{R}}{2}\,\overline{F(|\phi|^2)\,\phi}(u,\omega)=\sum_{n=0}^{\infty}\,
 \int\limits_{-\infty}^{\infty}
 \,F_n(u)\,F_n(u')\sqrt{ \omega^{1/2} R+2n+1}\,\overline{\phi}(u',\omega)\,d u',
 \end{eqnarray}


 or taking into account that $$(-\partial_u^2+\frac{\omega}{R^2}\,u^2)F_n=
 \frac{\omega^{1/2}}{R}\,F_n, $$
 in the equivalent form
 
\begin{eqnarray}\label{chane}
\left(\sqrt{-\partial_u^2+(1+\frac{u^2}{R^2})\,\omega}-\frac{\nu}{2}\right)\,
\bar{\phi}(u,\omega)+
\frac{\nu\,A}{2}\overline{|\phi|^2\,\phi}(u,\omega)=0.
 \end{eqnarray}
 Thus, eliminating the waves in the linear medium in which the waveguide is embedded and 
 applying  the inverse Fourier transformation with respect to Eq. (\ref{fourier}), leads to the
 following equation for the waveguide function:
 \begin{eqnarray}\label{wgf}\phi(u,t)\equiv \psi(u,R,t)\end{eqnarray}
 \begin{eqnarray}\label{chanpeq}
\left(\sqrt{-\partial_u^2-i\,(1+\frac{u^2}{R^2})\, \partial_t}-\frac{\nu}{2}\right)\,
\phi(u,t)+
\frac{\nu\,A}{2}|\phi(u,t)|^2\,\phi(u,t)=0.
 \end{eqnarray}
 Thus, the dynamics of the system is described by the pseudo-differential or, in other words,
 the nonlocal,    in time and space, equation. 
The nonlocal character of  the nonlinear waveguide dynamics
 is due to the existence  of two paths for the excitation energy transfer: directly  along 
 the waveguide and via the linear medium in which the waveguide is embedded.
 
 \subsection{Stationary states}
 
 We are interested here in the stationary solutions of Eq. (\ref{chanpeq}) of the form
 \begin{eqnarray}\label{stsol}\phi(u,t)=\Phi(u)\,e^{i\,\lambda^2\,t}.\end{eqnarray}
 Here the shape function $\Phi(u)$ satisfies the equation
 \begin{eqnarray}\label{chansteq}
\left(\sqrt{-\partial_u^2+(1+\frac{u^2}{R^2})\,\lambda^2}-\frac{\nu}{2}\right)\,
\Phi+
\frac{\nu\,A}{2}|\Phi|^2\,\Phi=0.
 \end{eqnarray}

Let us consider in more detail  the case of repulsive excitations ($A>0$). This case is the most 
interesting from the point of view of the interplay between  the nonlinearity and curvature, because, 
for the straight waveguide
($R\rightarrow \infty$),  Eq. (\ref{chansteq})  has no localized solutions. 

When the  nonlinearity parameter $A$ is 
sufficiently large, one can use the so-called 
Thomas-Fermi approximation \cite{thomasfermi} 
in which one neglects the term $\partial_u^2$ 
in  Eq. (\ref{chane}), and one finds a density profile
\begin{eqnarray}\label{thf}
|\phi(u)|^2=\frac{2\lambda}{\nu A }\left(\frac{\nu}{2\lambda}-\sqrt{1+u^2/R^2}\right),~~~~-
u_0\leq u\leq u_0,\nonumber\\
|\phi(u)|^2=0,~~~|u|\geq u_0,\end{eqnarray}
where $u_0=R\sqrt{\frac{\nu^2}{4\lambda^2}-1}.$ 

By using Eqs. (\ref{stsol}) and (\ref{eqparc}), 
it is straighforward to show that in terms of the channel wave  function
 (\ref{chanwf})
 the Hamiltonian (\ref{ham}) and 
 the norm (\ref{norm}) may be written as
 \begin{eqnarray}\label{hammod}
 H=-\frac{1}{4}\,\nu \,A\,R\,
 \int\limits_{-\infty}^{\infty}
 \,\phi^4(u)\,d u-\frac{\lambda^2}{2}\,N,
 \end{eqnarray}
 \begin{eqnarray}\label{normmod}
 N=-\frac{1}{2}\,\nu \,A\,R\,\frac{\partial}{\partial \lambda^2}
 \int\limits_{-\infty}^{\infty}
 \,\phi^4(u)\,d u.
 \end{eqnarray}

Inserting Eq. (\ref{thf}) into Eqs (\ref{hammod}) and  
(\ref{normmod}),  we obtain the following expressions for the 
energy   and the norm of the 
nonlinear excitations:
 \begin{eqnarray}H=-\frac{\nu R}{12 A}\,\{\left(5+\frac{16\lambda}{\nu^2}\right)\sqrt{\frac{\nu^2}{4\lambda^2}-1}-
-\frac{18 \lambda}{\nu}\, {\rm arcsinh} \left(\sqrt{\frac{\nu^2}{4\lambda^2}-1}\right)\}.\label{thfham}\end{eqnarray}
\begin{eqnarray}\label{thfnorm}
N=
\frac{R}{\lambda^2 A}\,\{\frac{\nu}{2\lambda}\sqrt{\frac{\nu^2}{4\lambda^2}-1}-
{\rm arcsinh} \left(\sqrt{\frac{\nu^2}{4\lambda^2}-1}\right)\},\end{eqnarray}

From  Eqs (\ref{thfham}) and  (\ref{thfnorm}) one can obtain  that 
for $N \nu A/R \,\gg  \,1$

$$H\approx -\frac{5}{6}\,
\left(\frac{\nu^2\,R}{A}\right)^{2/3}\,\left(\frac{N}{2}\right)^{1/3}, $$

while for $N \nu A/R\,\ll  \,1$

$$H\approx -\frac{\nu^2}{4}\,N +\frac{3}{20}\nu^{8/3}\,
\left(\frac{3 A }{4 R}\right)^{2/3}\,N^{5/3}.$$


\subsection{Non-stationary dynamics}

When $ R\rightarrow\infty$   Eq. (\ref{chanpeq}) assumes the form 
\begin{eqnarray}\label{chanstraight}
\left(\sqrt{-\partial_u^2-i\, \partial_t}-\frac{\nu}{2}\right)\,
\phi(u,t)+
\frac{\nu\,A}{2}|\phi(u,t)|^2\,\phi(u,t)=0,
 \end{eqnarray}
which is the nonlinear Hilbert-NLS equation introduced in Ref. \cite{krp}.  

In the limit of weak nonlinearity ($A\ll 1$)  and small curvature ($R\gg 1$)  one can significantly simplify 
Eq. (\ref{chanpeq})  by using the Ansatz
\begin{eqnarray}\label{ans}
\phi(u,t)=\exp\left(i \,t\,\nu^2/4 \right)\,\varphi(u,t)\end{eqnarray} and assuming that $\varphi(u,t)$ depends slowly on $u$ and $t$.
Inserting Eq. (\ref{ans}) into Eq. (\ref{chanpeq}), we obtain
\begin{eqnarray}\label{chans}
\left(\sqrt{-\partial_u^2+\,(1+\frac{u^2}{R^2})\,(\frac{\nu^2}{4} -i\,\partial_t)}-\frac{\nu}{2}\right)\,
\varphi(u,t)+
\frac{\epsilon\,A}{2}|\varphi(u,t)|^2\,\varphi(u,t)=0.
 \end{eqnarray}
 Considering the scaling
 \begin{eqnarray}\label{ord}
 t=\tilde{t}\,\epsilon^{-2}, ~~~u=\tilde{u}\,\epsilon^{-1},~~~R= \tilde{R}\,\epsilon^{-2}, ~~~A=\tilde{A} \,\epsilon^{2}\end{eqnarray}
 for $~\epsilon\rightarrow 0,~~$   Eq. (\ref{chans}) reduces to
\begin{eqnarray}\label{chan_appr}
i\,\frac{\partial}{\partial \tilde{t}}\varphi=-\frac{\partial^2}{\partial \tilde{u}^2}\,\varphi+\frac{\nu^2\tilde{u}^2}{\tilde{R}^2}\,\varphi+
\frac{\nu^2\tilde{A}}{2}\,|\varphi|^2\,\varphi.
 \end{eqnarray}
 Thus  the behaviour of nonlinear excitations in a  curved waveguide is equivalent to the behaviour of 
 nonlinear excitations  in  the parabolic potential whose curvature coincides with the 
 curvature of the waveguide. In the linear case
$A=0$ the corresponding eigenvalue problem    gives the 
eigenvalues presented in   Eq. (\ref{Rlarge}). Furthermore, it is interesting
that this reduction gives rise to an effective one-dimensional 
Gross-Pitaevskii (GP) equation, analogous to the one used in the
study of Bose-Einstein condensates in cigar-shaped traps \cite{GPE1d}.
While we do not pursue this analogy in detail in the present 
proof-of-principle paper, we note that it would naturally be of
interest to examine excitations known in the context of the GP
equation, such as e.g., dark solitons \cite{dark} and their  dynamics,
in the present setting.

\section{Numerical Results}

\begin{figure}
\centerline{\hbox{
\psfig{figure=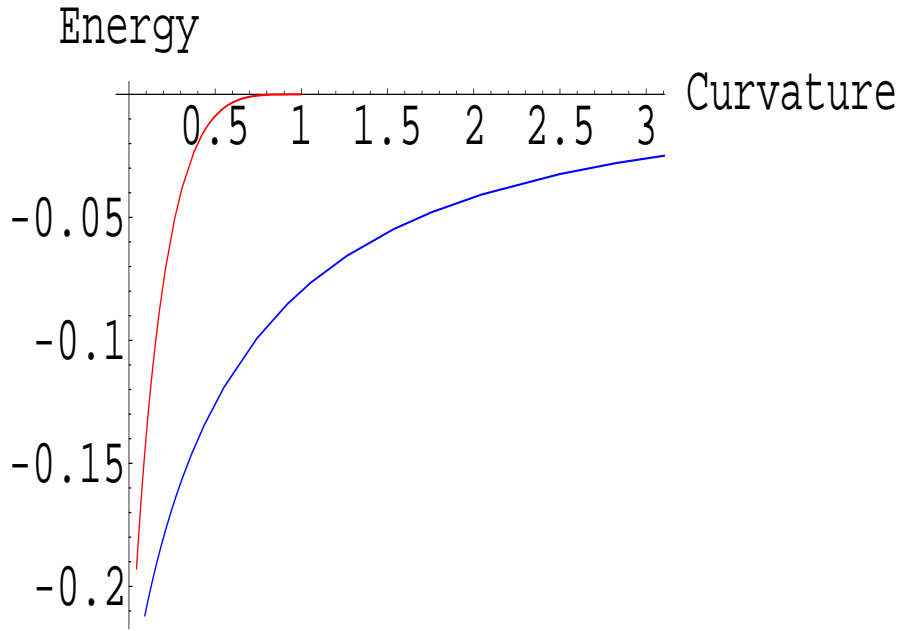,width=120mm,height=120mm,angle=0}}}
\vspace{3mm}
\caption{Bound state energy ($-\lambda_n^2$) {\it vs} curvature ($\kappa=1/R$): 
$n=0$ (blue line),  $n=1$ (red line). The dependence of the energy as
a function of the curvature is obtained through the solution of Eq. 
(\ref{lambda}).}
\label{fig:energy_vs_curv}
\end{figure}

\begin{figure}
\centerline{\hbox{
\psfig{figure=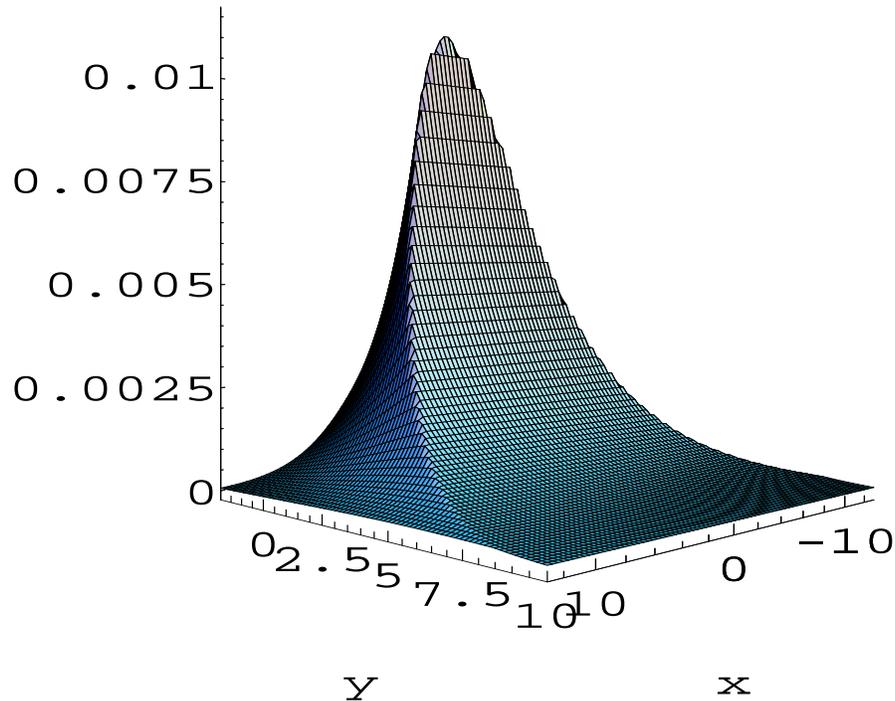,width=120mm,height=120mm,angle=0}}}
\vspace{3mm}
\caption{The wave function of the lowest bound state of the linear problem 
($A=0$) is shown in the figure for $R=10$, $\nu=1$
(obtained through Eq. (\ref{phi0})).}
\label{fig:wafefunc}
\end{figure}

We start by demonstrating the results of the linear case of
Eqs. (\ref{lambda}) and (\ref{phi0}). The eigenvalue (energy) of
the linear case as a function of curvature is shown in Fig. 
\ref{fig:energy_vs_curv}, while the lowest energy, bound state 
wavefunction of the linear problem is given in Fig. \ref{fig:wafefunc}.

\begin{figure}
\centerline{\hbox{
\psfig{figure=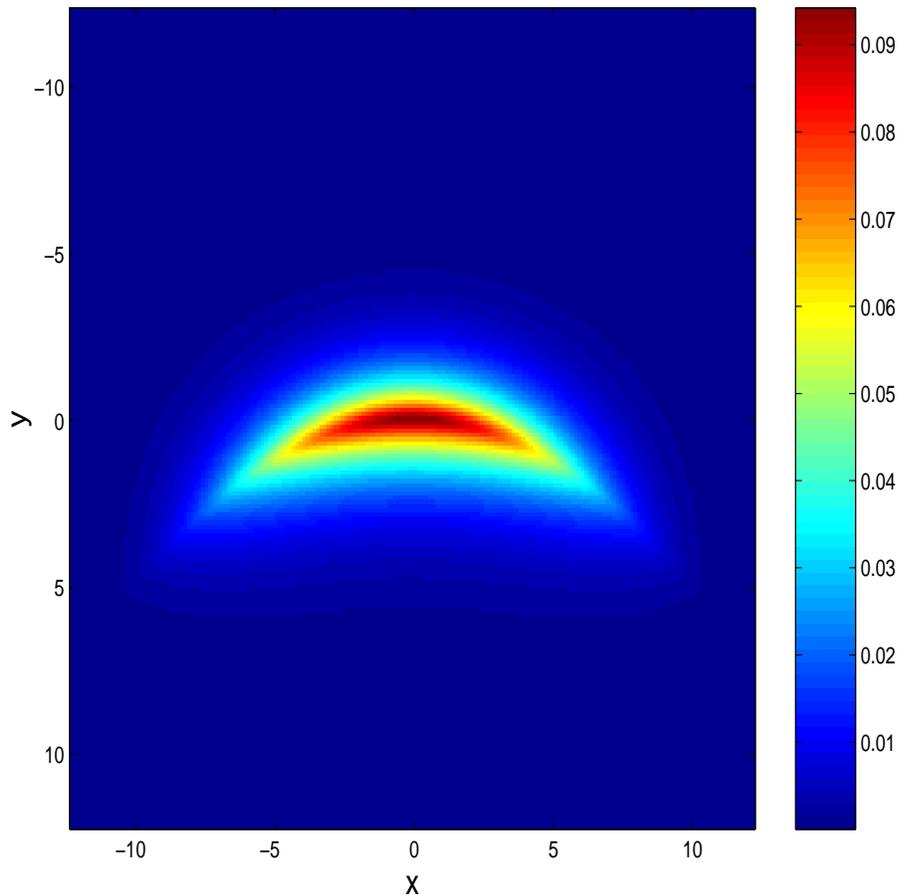,width=120mm,height=120mm,angle=0}}}
\vspace{3mm}
\caption{Contour plot of the solution of the partial differential
equation of Eq. (\ref{eqpar}) with $f(x)=x^2/(2 R)$, $R=10$, $\nu=A=1$
and initial condition given by the linear profile of Eq. (\ref{phi0}).}
\label{yfig3}
\end{figure}

In order to demonstrate that this linear bound state persists
in the nonlinear limit we have performed full dynamical evolution
simulations of Eq. (\ref{eqpar}), with an initial condition
of the form expressed in Eq. (\ref{phi0}), and demonstrated
in Fig. \ref{fig:wafefunc}. We note in passing that
similar results have been obtained with other initial conditions
such as e.g., a Thomas-Fermi initial profile of the form of Eq.
(\ref{thf}). 
In particular, we show typical 
numerical simulation results in Fig. \ref{yfig3} for $R=10$,
$\nu=A=1$. Notice that the $\delta$ function was represented
as 
\begin{eqnarray}
\delta(s)=\sqrt{\frac{1}{\pi \epsilon}} \exp \left(-\frac{s^2}{2 \epsilon}\right)
\label{delta1}
\end{eqnarray}
with $\epsilon=0.05$. 
The contour plot of Fig. \ref{yfig3} shows the result after a numerical 
evolution of 100 time units of Eq. (\ref{eqpar}).
The dynamical development indicates that after an initial transient
the original linear profile slightly reshapes itself into the
nonlinear solution depicted in Fig. \ref{yfig3}. In the process,
some radiation waves (``phonons'') are shed, that are absorbed by
the absorbing boundary conditions used in a layer close to the
the end of the domain (our computational box is of size $25\times25$).

\begin{figure}
\centerline{\hbox{
\psfig{figure=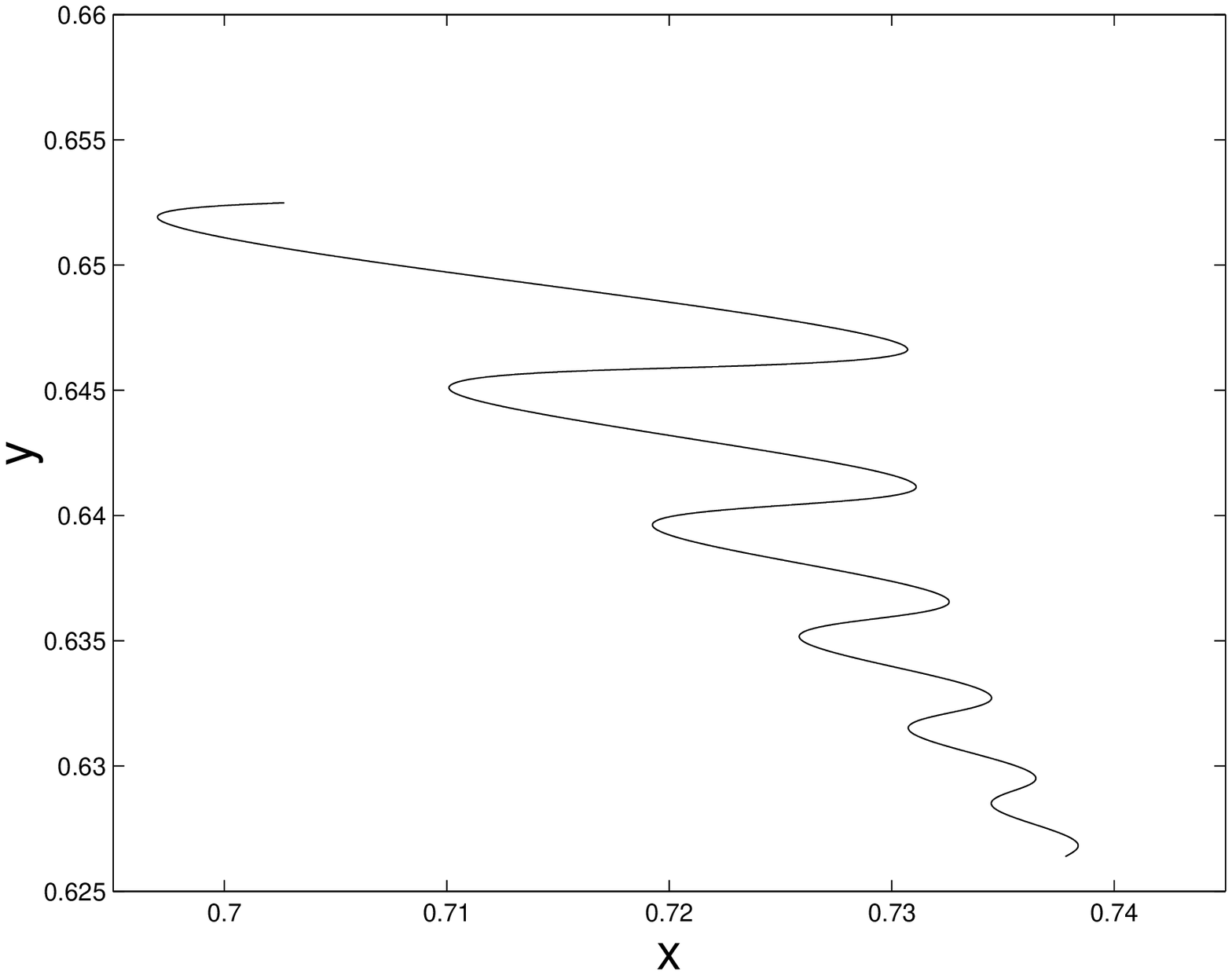,width=60mm,height=60mm,angle=0}}}
\centerline{\hbox{
\psfig{figure=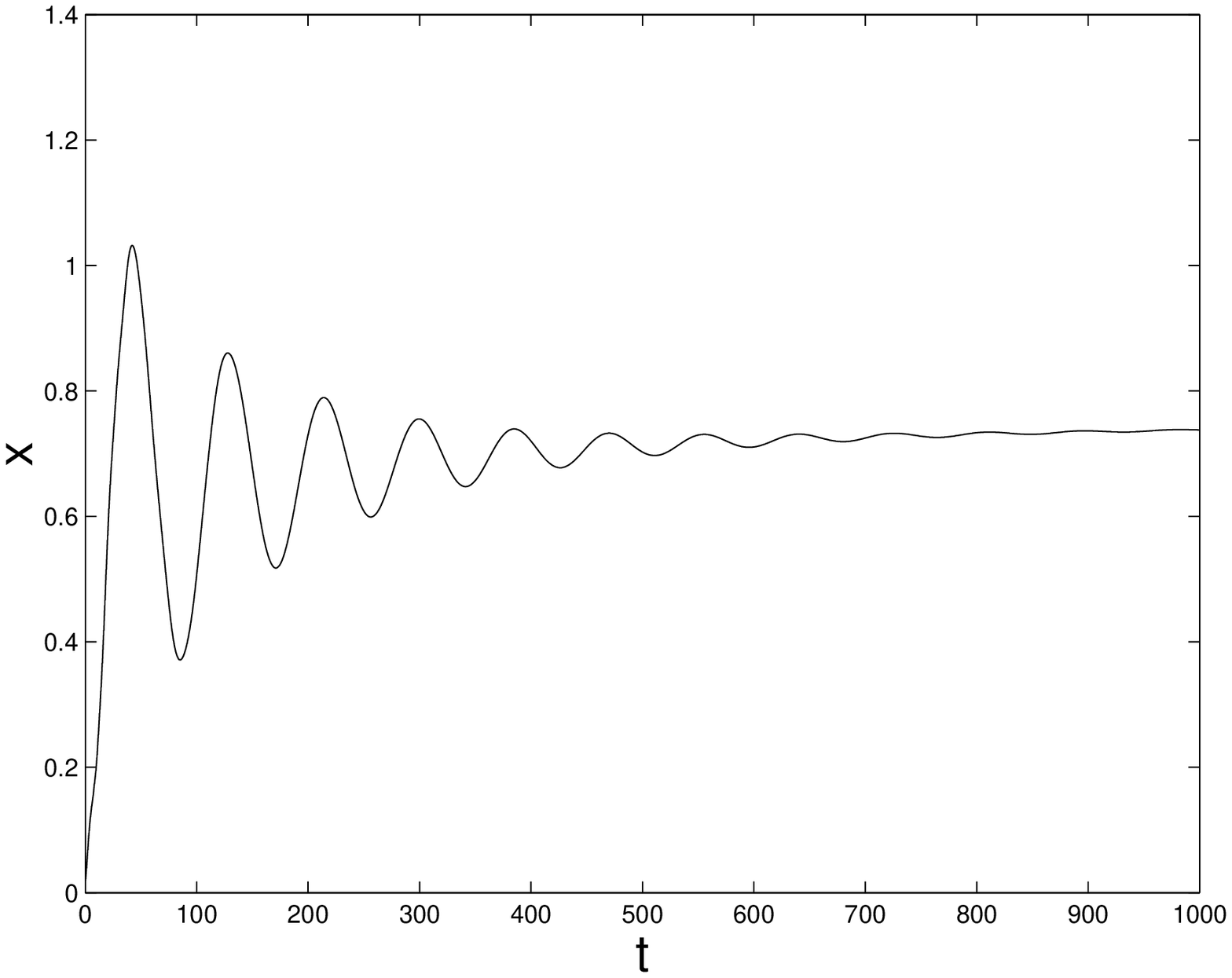,width=60mm,height=60mm,angle=0}}}
\centerline{\hbox{
\psfig{figure=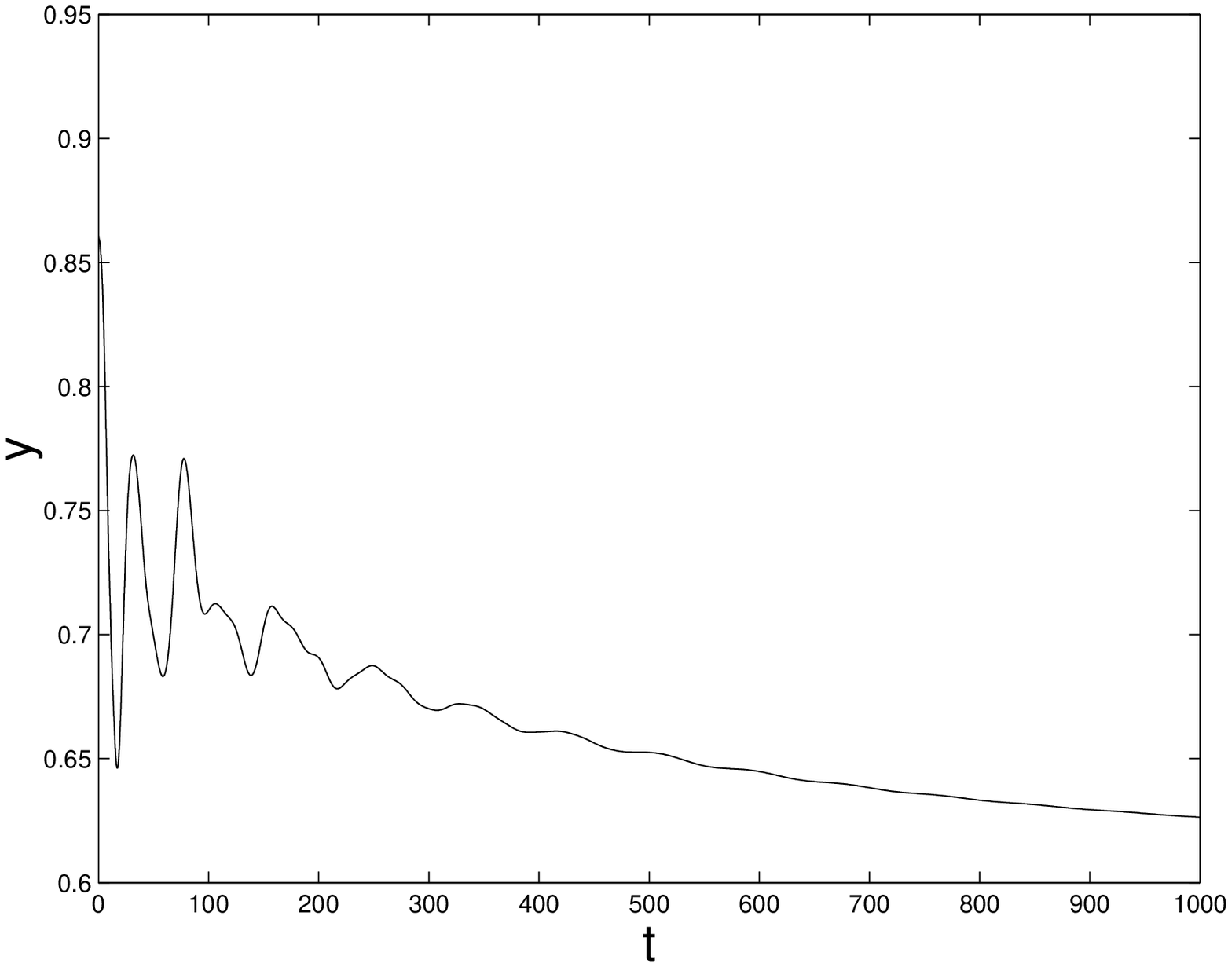,width=60mm,height=60mm,angle=0}}}
\vspace{3mm}
\caption{The top panel shows, for times between $t=500$
and $t=1000$, the time evolution of the center of mass
in the $x-y$ plane (the motion proceeds from top left to
bottom right as time evolves). The middle panel shows
the time evolution of the $x$-position of the center
of mass of the nonlinear excitation, while the bottom
panel the slow evolution of the $y$-position of
the center of mass.}
\label{yfig4}
\end{figure}

Beyond the proof-of-principle simulations for various initial
conditions theoretically derived in sections III and IV, we also
attempted to examine the dynamics of the nonlinear excitations
of the channel. This was done using the following numerical
protocol: after obtaining a quasi-relaxed nonlinear localized
mode for the channel of the form $y=x^2/(2 R)$, we moved the
channel to a new position, namely $y=(x-1)^2/(2 R)$. Notice that
similar in spirit experiments have recently been carried in Bose-Einstein
condensates \cite{smerzi}, where the magnetic trap confining the
condensate is displaced to a new position and the ensuing dynamics
of the condensate are observed. The position of the center of
mass of the initial 
condition profile was approximately obtained (using a trapezoidal
approximation to the relevant two-dimensional integrals) as 
$(x(t=0),y(t=0))=(0.178,0.862)$. The new bottom of the channel
(hence the point to which the center of mass should approach)
in this case is $(x,y)=(1,0)$. In Fig. \ref{yfig4}, the
process of relaxation to this new equilibrium is shown 
as a function of time for a very long dynamical simulation
of $t$ up to $1000$ time units. In this run, we observe
(after an initial transient) a slow relaxation towards the
new minimum of the potential well. Notice that despite the
{\it Hamiltonian} nature of the model, the excitation of
an ``internal mode'' of the nonlinear wave \cite{internal}
can be dissipated due to mechanisms of coupling to 
extended wave, phonon modes, such as the ones
reported in \cite{dissip}.  

\section{Summary}

In summary, we have shown that:
\begin{itemize}
\item In two-dimensional media with a curved {\it infinitely thin}   waveguide 
(quantum channel) 
there exist bound states for linear and nonlinear self-interacting excitations;
\item The finite curvature of the waveguide provides a stabilizing effect  
on otherwise unstable localized states of repelling  excitations;
\item  The binding energy of both linear and nonlinear  localized excitations decreases when the curvature of the waveguide increases.
\item Such linear bound states as the ones found here persist in the
nonlinear dynamical problem as localized excitations. These have been
found to be robust for different initial conditions and are centered
at the minimum (point of largest curvature) of the parabola. When,
the mode is initialized away from this minimum, it slowly relaxes
to it.  
\end{itemize}

While our motivation was originally provided by the embedding of a waveguide
in a two-dimensional photonic crystal (as well as from more general geometric
considerations), interesting analogies have arisen through our investigation,
that warrant further study. In particular, we note the analogy
in the weakly nonlinear regime of the equation for the waveguide/quantum
channel with that of the Bose-Einstein 
condensate behavior in the presence of a magnetic trap in atomic physics. 
Another topic worthy of further investigation is a more detailed 
numerical study of the stability of the nonlinear localized
modes we have identified. Finally, another direction that could
be of further interest is the examination of a case
of a finite (rather than infinitesimal) width channel [which can
be computational achieved e.g., by allowing the parameter $\epsilon$
of Eq. (\ref{delta1}) to vary towards larger values]. The examination
of thresholds for  genuinely two-dimensional instabilities, such as 
e.g., the transverse or the snaking instability, would be of
particular interest within the latter context.

Such studies are currently in progress and will be reported in
future publications. 

\section*{Acknowledgments}
Yu.B.G.  thanks  the Informatics and  Mathematical Modelling, Technical
University of Denmark for a Guest professorship.
 Yu.B.G acknowledges support from Deutsche Zentrum f{\"u}r Luft- und Raumfart
 e.V., Internationales B{\"u}ro des Bundesministeriums  f{\"u}r Forschung
 und Technologie, Bonn, in the frame of a bilateral scientific cooperation 
 between Ukraine and Germany, project No. UKR-02/011.
This work was supported by a UMass FRG, NSF-DMS-0204585 and the
Eppley Foundation for Research (PGK)
 
\appendix{}
\section{}
  To find  an expression  for the Green's function $G(u,u';v,v')$ we expand this function in terms
  of of the eigenfunctions of the equation
  
\begin{eqnarray}\label{eqF}
\frac{d^2 F}{d\sigma^2}-\sigma^2\,F=-\mu F,
\end{eqnarray}

under the boundary conditions
\begin{eqnarray}\label{bcF}F \rightarrow  0,~~~~~u\rightarrow \pm\infty.
\end{eqnarray}
 
 Eq. (\ref{eqF}) under the boundary conditions (\ref{bcF}) has the solutions
 \begin{eqnarray}\label{solF}
 F_n(u)=a_n\,e^{-\sigma^2/2}\,H_n\left(\sigma\right),~~~n=0,1,2,...
 \end{eqnarray}
 ($\sigma=u\,\sqrt{\lambda/R}$),  which correspond to the eigenvalues \begin{eqnarray}\label{eig}  \mu= 2n+1.
 \end{eqnarray} 

Here  
 $H_n\left(z\right)$ is the Hermite polynomial \cite{abr} and $a_n=\left(\lambda/R\pi\right)^{1/4}\,
 \left(2^n\,n!\right)^{-1/2}$ is the normalization constant. In terms of the functions (\ref{solF}), the
 Green's function has the form
 \begin{eqnarray}\label{expgr}G(u,u';v,v')=\sum_{n=0}^{\infty}F_n(u)\,F_n(u')\,g_n(v,v').\end{eqnarray}
 Inserting Eq. (\ref{expgr}) into Eq. (\ref{eqgreen}) for the coefficients $g_n(v,v')$ we obtain  the equation  
 
 \begin{eqnarray}\label{equG}
 \frac{d^2 g_n}{d v^2}-\frac{\lambda^2}{R^2}v^2\,g_n- \frac{\lambda}{R}(2n+1)\,g_n=\delta(v-v'), \end{eqnarray}
  under the condition
 \begin{eqnarray}\label{bcG}g_n \rightarrow  0,~~~~~v\rightarrow \infty.
 \end{eqnarray} The line $v=0$, which is the
 $y>R/2$ of the ordinate axis,  is used as the cut \cite{morse}. Following the
 usual machinery (see e.g. \cite{morse}), to avoid 
 discontinuities in the solution along the cut we shall require that
 \begin{eqnarray}
\frac{d g_n}{d v}=0~~~ at ~~v =0 ~~~~~when~ n= 2 m,\nonumber\\
 g_n(0)=0 ~~~when ~~n=2 m+1 ~~(m=0,1,2,..). \label{cut} \end{eqnarray}

The solution of the equation (\ref{equG}) with regard to
 the boundary conditions (\ref{bcG}) and (\ref{cut}) has the form
 \begin{eqnarray}\label{solG}
 g_{n}= \sqrt{\frac{\pi\,R}{4\lambda}}\,
\{V(n+\frac{1}{2}, v\,\sqrt{2\lambda/R})\,U(n+\frac{1}{2}, v'\,\sqrt{2\lambda/R})\,\theta(v'-v)+\nonumber\\
U(n+\frac{1}{2}, v\,\sqrt{2\lambda/R})\,V(n+\frac{1}{2}, v\,\sqrt{2\lambda/R})\,\theta(v-v')\}
 \end{eqnarray}
 where $U(a,x)$ and $V(a,x)$ are the Weber parabolic 
 cylinder functions \cite{abr} and $ \theta(v)$ is the Heaviside step-function. 
 Another equivalent representaion in terms of Whittaker functions \cite{abr}
 is
 \begin{eqnarray}\label{whit}
 U(n+\frac{1}{2},x)=2^{-n/2-1/4}\,
 W\left(-\frac{n}{2}-\frac{1}{4},-\frac{1}{4},\frac{1}{2}\,x^2\right),\nonumber\\
 V(n+\frac{1}{2},x)=\frac{2^{-n/2+3/4}\,n!}{\sqrt{\pi\,x}\,\Gamma(\frac{n}{2}+1)}\,
 M\left(-\frac{n}{2}-\frac{1}{4},\frac{1}{4},\frac{1}{2}\,x^2\right).
 \end{eqnarray}


\begin{thebibliography}{99}
\bibitem{christ} {\emph Nonlinear Science at the Dawn of the 21st Century}, Eds P.L. Christiansen, M.P. S{\o}rensen,
A.C. Scott, Lecture Notes in Physics (Springer,Berlin, Heidelberg, New York,2000);
{\emph Localization and Energy transfer in Nonlinear Systems},
Eds L. V{\'a}zquez, R.S. Mackay, M.P.Zorzano (World Scientific, Singapore, 2003).
\bibitem{souk} {\emph Photonic Crystals and Light Localization in the 21st
Century}, C. M. Soukoulis (ed.), NATO Science Series C{\bf 563}, 
(Kluwer Academic 
Dordrecht, Boston, London, 2001).

\bibitem{mekis1} A. Mekis, J.C. Chen, I. Kurland, S. Fan, P.R. Villeneuve, and 
J.D. Joannopoulos, 
Phys. Rev. Lett. {\bf 77}, 3787 (1996).

\bibitem{noda} S. Noda, A. Alongkarn, and 
M. Imada, Nature (London), {\bf 407}, 
608 (2000).

\bibitem{kkt} Yu.S. Kivshar, P.G. Kevrekidis and S. Takeno,
Phys. Lett. A, {\bf 307}, 287 (2003).


\bibitem{wees} B.J. van Wees, H. van Houten, C.W.J. Beenakker, 
J.G. Williamson, L.P. Kouwenhoven, 
D. van der Marel, and C.T. Foxon, Phys. Rev. Lett. {\bf 60}, 848 (1988).


\bibitem{reed} {\emph Nanostructure Physics and Fabrication}, Eds M.A. Reed and W.P. Kirks,
(Academic Press, New York,1989).
\bibitem{ismail} K. Ismail, S. Washburn, and K. Y. Lee, Appl. Phys. Lett. 
{\bf 59}, 1998 (1991).
\bibitem{Pulwey01}
R. Pulwey, M. Rahm, J. Biberger, and D. Weiss, IEEE Transactions on magnetics 
{\bf 37,} 2076
  (2001).
 \bibitem{shinjo}T. Shinjo, T. Okuno, R. Hassdorf, K. Shigeto, and T. Ono, Science, 
 {\bf 289}, 930 (2000).
 \bibitem{klaeui} M. Klaui, C. A. F. Vaz, J. Rothman, 
 J. A. C. Bland, W. Wernsdorfer, G. Faini, and E. Cambril, 
 Phys. Rev. Lett. {\bf 90}, 097202 (2003).
 \bibitem{schult} R.L. Schult, D.G. Ravenhall and H.W. Wyld, Phys. Rev. B{\bf 39}, 5476 (1989).
 \bibitem{berggren} K.F. Berggren and Z.L. Ji, Phys. Rev. B{\bf 43}, 4760 (1991).
 \bibitem{andrews} M. Andrews and C.M. Savage , Phys. Rev. A {\bf 50}, 4535 (1994).
 \bibitem{exner} P. Exner and P. Seba, J. Math. Phys. {\bf 30}, 2574 (1989).
\bibitem{gold} J. Goldstone and R.L Jaffe, Phys. Rev. B {\bf 45},14100 (1992).
\bibitem{vakh} Yu. B. Gaididei and O.O. Vakhnenko, J. Phys.: Condens. Matter {\bf 6}, 32229 (1994);
O.O. Vakhnenko, Phys. Rev. B{\bf 52}, 17386 (1995).
\bibitem{exner01} P. Exner and T. Ichinose, J. Phys. A {\bf 34}, 1439 (2001).
\bibitem{costa} R.C.T. da Costa, Phys. Rev. A {\bf 23}, 1982 (1981).
\bibitem{jaffe} P.C. Schuster and R.L. Jaffe, e-print hep-th/0302216.
\bibitem{exnkr} P. Exner and D. Krejcirik, J. Phys. A: Math. Gen. {\bf 34}, 5969 (2001).
\bibitem{encinosa} M. Encinosa and B. Etemadi, Phys. Rev. A {\bf 58}, 77 (1998);
 M. Encinosa and L. Mott,  Phys. Rev. A {\bf 68}, 014102 (2003).
\bibitem{harayama} T. Harayama, P. Davis and K. S. Ikeda, Phys. Rev. Lett. 
{\bf 82}, 3803 (1999).
\bibitem{ustinov}  A. Wallraff, A.V. Ustinov, V.V. Kurin, I.A. Shereshevsky, 
and N.K. Vdovicheva, Phys. Rev. Lett. {\bf 84}, 151 (2000).
\bibitem{mingkivsh} S.F. Mingaleev, Yu.S. Kivshar, and R.A. Sammut, Phys. Rev. E {\bf 62}, 5777 (2000); S.F. Mingaleev and Yuri S. Kivshar 
                    Phys. Rev. Lett. {\bf 86}, 5474 (2001).


\bibitem {curve}Yu. B. Gaididei, S.F. Mingaleev and P.L. Christiansen, 
Phys. Rev. E {\bf62} R53 (2000); J. Phys.:Condens. Matter {\bf 13},1 (2001).
\bibitem{leanhardt} A.E. Leanhardt, A.P. Chikkatur, D. Kielpinski, Y. Shin, T.L. Gustavson, W. Ketterle, 
and D.E. Pritchard, Phys. Rev. Lett. {\bf 89}, 040401 (2002).
\bibitem{encyklop} {\emph Encylopaedia of Mathematics}, vol. 3 (Kluwer Academic Publishers, Dordrecht, Boston, London, 1989),
p.  159.
\bibitem{djones} D. Jones, {\it Generalized Functions} (McGraw-Hill 
Publishing Company, London 1966).

\bibitem{abr} M.~Abramowitz and I.~Stegun, {\em Handbook of 
Mathematical Functions} (Dover Publications, Inc., New York, 1972).

\bibitem{thomasfermi} see e.g., 
F. Dalfovo, S. Giorgini, L. P. Pitaevskii, and S.
Stringari, 
Rev. Mod. Phys. {\bf 71}, 463
(1999). 

\bibitem{krp} Yu. B. Gaididei, P.L. Christiansen,  K. {\O}. Rasmussen, 
and M. Johansson, Phys. Rev. B {\bf 55} R13365 (1997).

\bibitem{GPE1d} V.M.\ P\'{e}rez-Garc\'{\i}a, H.\ Michinel and H.\ Herrero,
Phys.\ Rev.\ A \textbf{57}, 3837 (1998); L.\ Salasnich, 
A.\ Parola and L.\ Reatto,
Phys.\ Rev.\ A \textbf{65}, 043614 (2002); Y.B. Band, I. Towers, and B.A.
Malomed, Phys.\ Rev.\ A 67, 023602 (2003).

\bibitem{dark} W.P. Reinhardt and C.W. Clark, J. Phys. B {\bf {30}}, L785
(1997); Th. Busch and J.R. Anglin, Phys. Rev. Lett. {\bf {84}}, 2298 (2000);
D.J. Frantzeskakis {\it et al.}, 
G. Theocharis, F.K. Diakonos, P. Schmelcher, and Yu.S. Kivshar, 
Phys. Rev. A {\bf 66}, 053608 (2002);
P. G. Kevrekidis, R. Carretero-Gonz{\'a}lez, 
G. Theocharis, D. J. Frantzeskakis, and B. A. Malomed,
Phys. Rev. A {\bf 68}, 035602 (2003).

\bibitem{smerzi} A. Smerzi, A. Trombettoni, P.G. Kevrekidis and A.R. Bishop,
\newblock Phys. Rev. Lett. {\bf 89}, 170402 (2002); 
F.S. Cataliotti, L. Fallani, F. Ferlaino, C. Fort, P. Maddaloni and
M. Inguscio, New. J. Phys. {\bf 5}, 71 (2003). 

\bibitem{internal} 
Yu.S. Kivshar, D.E. Pelinovsky, Thierry Cretegny, and Michel Peyrard,
Phys. Rev. Lett. {\bf 80}, 5032 (1998); P.G. Kevrekidis and C.K.R.T. Jones,
Phys. Rev. E {\bf 61}, 3114 (2000).

\bibitem{dissip} M. Johansson and S. Aubry, Phys. Rev. E {\bf 61}, 5864
(2000);
P.G. Kevrekidis and M.I. Weinstein,
Physica D {\bf 142}, 113 (2000); P.G. Kevrekidis and M.I. Weinstein,
Math. Comp. Simul. {\bf 62}, 65 (2003).


\bibitem{morse} P.M. Morse and H. Feshbach {\em Methods of Theoretical Physics}
(McGraw-Hill Publishing Company, New York 1953).

\end{thebibliography}
\end{document}